%% file: main_springer.tex
\documentclass[runningheads]{llncs}
\usepackage[T1]{fontenc}
\usepackage{graphicx}
\input{packages_springer}
\makeatletter
\def\@fnsymbol#1{\ensuremath{\ifcase#1\or \dagger\or \ddagger\or
\mathsection\or \mathparagraph\or \|\or **\or \dagger\dagger
\or \ddagger\ddagger \else\@ctrerr\fi}}
\makeatother
\begin{document}
\title{A Genetic Algorithm-based Framework for \\ Learning Statistical Power Manifold \thanks{\scriptsize{Accepted for an oral presentation at the International Meeting of Psychometric Society (IMPS) 2022, Bologna, Italy.}}}
\titlerunning{Genetic Algorithm-based Learning of Statistical Power}
%
\author{Abhishek K. Umrawal \inst{1,2} 
\and
Sean P. Lane \inst{1,3} 
\and
Erin P. Hennes \inst{1,3} 
}
\authorrunning{A. Umrawal et al.}
%
\institute{Purdue University, West Lafayette, IN 47906, USA \\ 
\email{\{aumrawal, lane84, ehennes\}@purdue.edu} \and 
University of Maryland, Baltimore County, Baltimore, MD 21250, USA \\
\email{\{aumrawal\}@umbc.edu} \\ \and
University of Missouri, Columbia, MO 65211, USA \\
\email{\{lanesp, ehennes\}@missouri.edu}}

\maketitle              
\input{sections/abstract_springer}
\input{sections/introduction}
\input{sections/preliminaries}
\input{sections/methodology}
\input{sections/experiments}
\input{sections/conclusion}

\bibliographystyle{apalike}
\bibliography{refs}

\end{document}

%% file: packages_springer.tex
\usepackage{hyperref}
\usepackage{amsmath}
\usepackage{amsfonts}
\usepackage{amssymb}
\usepackage{cite}
\usepackage{natbib}
\usepackage{algorithm}
\usepackage{algpseudocode}
\usepackage{multirow}
\usepackage{multicol}
\usepackage{mathtools}

\usepackage[utf8]{inputenc}
\usepackage{pgfplots}
\usepgfplotslibrary{groupplots,dateplot}
\usetikzlibrary{patterns,shapes.arrows}
\pgfplotsset{compat=newest}
\pgfplotsset{scaled y ticks=false}
\usepackage{caption}

\usepackage{xcolor}

\usepackage{soul}

\usepackage{tikz}
\usetikzlibrary{shapes,arrows}
\tikzstyle{decision} = [diamond, draw, fill=blue!15, 
    text width=4em, text badly centered, node distance=3.5cm, inner sep=0pt]
\tikzstyle{block} = [rectangle, draw, fill=yellow!10,
    text width=12em, text centered, rounded corners, minimum height=2em]
\tikzstyle{line} = [draw, -latex']
\tikzstyle{cloud} = [draw, ellipse,fill=green!15, node distance=2cm,
    minimum height=2em]

%% file: sections/abstract_springer.tex
\begin{abstract}
Statistical power is a measure of the replicability of a categorical hypothesis test. Formally, it is the probability of detecting an effect, if there is a true effect present in the population. Hence, optimizing statistical power as a function of some parameters of a hypothesis test is desirable. However, for most hypothesis tests, the explicit functional form of statistical power for individual model parameters is unknown; but calculating power for a given set of values of those parameters is possible using simulated experiments. These simulated experiments are usually computationally expensive. Hence, developing the entire statistical power manifold using simulations can be very time-consuming. We propose a novel genetic algorithm-based framework for learning statistical power manifolds. For a multiple linear regression $F$-test, we show that the proposed algorithm/framework learns the statistical power manifold much faster as compared to a brute-force approach as the number of queries to the power oracle is significantly reduced. We also show that the quality of learning the manifold improves as the number of iterations increases for the genetic algorithm. Such tools are useful for evaluating statistical power trade-offs when researchers have little information regarding a priori  `best guesses'  of primary effect sizes of interest or how sampling variability in non-primary effects impacts power for primary ones.
\end{abstract}

\keywords{Statistical power \and Hypothesis testing \and Genetic algorithm \and Nearest neighbors.}

%% file: sections/introduction.tex
\section{Introduction} \label{sec:introduction}
\subsection{Motivation} 
Statistical power analysis is of great importance in empirical studies \citep{yang2022low,fraley2014n,cafri2010meta}. Statistical power is a measure of the goodness/strength of a hypothesis test. Formally, it is the probability of detecting an effect, if there is a true effect present to detect. For instance, if we use a test to conclude that a specific therapy or medicine is helpful in anxiety and stress alleviation then the power of the test tells us how confident we are about this insight. Hence, optimizing the statistical power as a function of some parameters of a hypothesis test is desirable. However, for most hypothesis tests, the explicit functional form of statistical power as a function of those parameters is unknown but calculating statistical power for a given set of values of those parameters is possible using simulated experiments. These simulated experiments are usually computationally expensive. Hence, developing the entire statistical power manifold using simulations can be very time-consuming. The main motivation of this paper is to develop a framework for learning statistical power while significantly reducing the cost of simulations.

\subsection{Literature Review}
\citet{bakker2012rules, bakker2016researchers, maxwell2004persistence, cohen1992things} point out that quite frequently the statistical power associated with empirical studies is so low that the conclusions drawn from those studies are highly unreliable. This happens primarily due to the lack of a formal statistical power analysis. Using simulations of statistical power \citet{bakker2012rules} shows that empirical studies use questionable research practices by using lower sample sizes with more trials. The results show that such practices can significantly overestimate statistical power. This makes the conclusions of the study about the effect size to be misleading as the reproducibility of the study is hampered due to the low statistical power. The questionable research practices include performing multiple trials with a very small sample size, using additional subjects before carrying out the analysis, and removal of outliers. \citet{bakker2012rules} emphasizes carrying out a formal power analysis for deciding the sample size to avoid such low statistical power. Recently, \citet{baker2021power} provides an online tool for drawing power contours to understand the effect of sample size and trials per participant on statistical power. The results provided in this paper demonstrate that changes to the sample size and number of trials lead to understanding how power regions of the power manifold. \citet{rast2014longitudinal} also demonstrates a powerful (inversely correlated) impact of sample size on both the effect size and the study design.  Recently, \citet{lane2018power} and  \citet{lane2019conducting} have provided simulation-based methods for formally conducting statistical power analysis. These simulation-based methods perform well in terms of power estimation but can be computationally expensive due to a high number of queries to the simulation-based power function oracle.

\subsection{Contribution} 
In this paper, we provide a novel genetic algorithm-based framework for learning the statistical power manifold in a time-efficient manner by significantly reducing the number of queries to the power function oracle. For a multiple linear regression $F$-test, we show that the proposed algorithm/framework learns the statistical power manifold much faster as compared to a brute-force approach as the number of queries to the power oracle is significantly reduced. We show that the quality of learning the manifold improves as the number of iterations increases for the genetic algorithm.

\subsection{Organization}     
The rest of the paper is organized as follows. We first discuss some preliminaries and formulate the problem. We next discuss the proposed methodology in detail with application to the multiple linear regression model. We next provide experiments to demonstrate the performance of our algorithm. Finally, we conclude the paper and discuss some potential future work.

%% file: sections/preliminaries.tex
\section{Preliminaries} \label{sec:preliminaries}

We start with some basic notations and preliminaries. Let $y_1,...,y_n$ be $n$ sample observations from some probability distribution $F$ with a $p$-dimensional parameter vector $\boldsymbol{\theta}$. A \textit{hypothesis test} $\phi$ is a function of the sample observations to accept or reject a specific hypothesis about the population parameter(s). The hypothesis of interest is called the \textit{null hypothesis} $(H_0)$ which is usually a hypothesis of no difference. The hypothesis that is alternative to the null hypothesis is called the \textit{alternative hypothesis} ($H_1$). A test can involve the following two types of errors.

\textit{Type I error} occurs when the test rejects the null hypothesis when it was true, i.e. the occurrence of false positives. The probability of Type I error is called the \textit{level of significance} which is same as the size of the critical region associated with the test. We denote the probability of Type I error as $\alpha$.
\begin{align}
    \alpha = \text{Prob}\left(\text{Reject } H_0 \text{ when it is true.}\right)
\end{align}
\textit{Type II error} occurs when the test accepts the null hypothesis when it was false, i.e. the occurrence of false negatives. We denote the probability of Type II error as $\gamma$ and $1-\gamma$ is defined as the \textit{power of the test}.
\begin{align}
    \gamma &= \text{Prob}(\text{Accept } H_0 \text{ when it is false.})\\
    \Rightarrow 1 - \gamma &= \text{Prob}(\text{Reject } H_0 \text{ when it is false.})
\end{align}
Hence, the power of the test is nothing but a measure of the test's ability to correctly detect the true positives.

In experimental studies, commonly the goal is to study some underlying effect/trait present in the population. We define \textit{effect size} ($\beta$) as a quantification of an effect present in the population calculated using some statistical measure. For instance, Pearson's correlation coefficient for the relationship between variables, Cohen's $d$ for the difference between groups, etc. Effect size is nothing but a function of the unknown population parameter(s). We can estimate effect size using the sample observations.

\textit{Statistical power analysis} involves studying the four variables defined above namely the effect size ($\beta$), sample size ($n$), level of significance ($\alpha$), and statistical power ($1-\gamma$). These variable are related to each other. For example, having a larger sample size makes it easier for the test to detect the false positives, and hence yields a higher statistical power. We are interested in estimating or optimizing one of the these four variables given the rest three. Most commonly, it is the statistical power that we want to optimize, i.e. how does the statistical power behaves given the other three variables viz. the effect size, sample size, and level of significance.

In experimental studies, the level of significance is usually predetermined as a fixed number, the sample size is given to be varying in a fixed range according to the budget constraint, and there is some expert/prior knowledge about the range in which the unknown effect size of interest may vary. There can be many questions surrounding statistical power analysis. Perhaps the most common question that arises around power analysis is that, what is the minimum sample size needed to attain a specific level of statistical power, given the magnitudes of the effects of interest and the level of significance. Hence, understanding the behavior of statistical power as a function of effect size, sample size, and level of significance is of great importance in experimental studies. One of the main challenges in doing so is that for most hypothesis tests, the explicit functional form of statistical power as a function of effect size, sample size, and level of significance is unknown. However, calculating statistical power for a given set of values of these variables is possible using simulated experiments. 

These simulated experiments are usually computationally expensive. Hence, developing the entire statistical power manifold using simulations can be very time-consuming. The problem of interest in this paper is to \textit{learn the statistical power manifold well in terms of power estimation in a time-efficient manner}.

%% file: sections/methodology.tex
\section{Methodology} \label{sec:methodology}

For the hypothesis test $\phi$ under consideration, let $\beta$ be the effect size. As discussed in the previous section, we know that $\beta$ is nothing but a function of the $p$-dimensional parameter vector $\boldsymbol{\theta}$ associated with the probability distribution $F$.

As discussed earlier, we know that in experimental studies the level of significance is usually predetermined as a fixed number, the sample size is given to be varying in a fixed range according to the budget constraint, and there is some expert/prior knowledge about the range in which the unknown effect size of interest may vary. Let $[\beta_l,\beta_u]$, $[\boldsymbol{\theta}_l,\boldsymbol{\theta}_u]$, and $[n_l,n_u]$ be the initial ranges of effect size, parameter vector, and sample size respectively. Note that, we only need either initial range for the effect size of the parameter vector.

Statistical power associated with a hypothesis is a function of the effect size $\beta$ (hence of $\boldsymbol{\theta}$), level of significance $\alpha$, and sample size $n$. As discussed earlier the level of significance is predetermined as a fixed number, we can say that power is a function of the effect size and sample size for a fixed given value of the level of significance.

Define the following vector.
\begin{align}
    \boldsymbol{c} := (\beta,n) \equiv (\boldsymbol{\theta},n) \equiv (\theta_1,...,\theta_p, n)
\end{align}

For simplicity and to avoid any ambiguity, we choose to work with $\boldsymbol{\theta}$ and not with $\beta$.

For the hypothesis test $\phi$ under consideration, statistical power is a function of the vector $\boldsymbol{c}$ for a given value of $\alpha$ denoted as follows.
\begin{align}
    1-\gamma := f_{\phi,\alpha}(\boldsymbol{c}) 
\end{align}

As discussed earlier, in most cases, we do not know the explicit algebraic form of $f(\cdot)$ to calculate power as a function of $\boldsymbol{c}$. However, calculating power for a specific choice of  $\boldsymbol{c}$ is possible using simulations. For our work, we assume that we have access to such a black box that takes $\boldsymbol{c}$ as input and returns $1-\gamma$ as output. We call this black box a \textit{power function value oracle}. A \textit{brute-force} way of learning/computing the power manifold (for parameters in the initial ranges) is to divide the ranges of the parameters into a high-dimensional grid, compute the power using the power function value oracle, and then plot the values. The drawback of this approach is the computational cost associated with a very large number of queries to the value oracle. Motivated by the idea of reducing the computational cost of learning the statistical power manifold while being able to do well in terms of power estimation, we propose a genetic algorithm-based framework.

Genetic algorithm \citep{goldberg1988genetic, holland1992genetic} is a meta-heuristic inspired by biological evolution based on Charles Darwin's theory of natural selection that belongs to the larger class of evolutionary algorithms. Genetic algorithms are commonly used to generate high-quality solutions to optimization and search problems by relying on biologically inspired operators such as mutation, crossover, and selection \citep{mitchell1998introduction}.

We next explain the steps involved in genetic algorithm in the context of our problem. 

We start with $N$ (a hyper-parameter) randomly chosen $\boldsymbol{c}$ vectors where entries inside the vector are chosen randomly from the respective ranges with some discretization. The discretization step size for parameters $\theta_1,...,\theta_p$ is usually fractional and for sample size is integral. Each of these $N$ random vectors is called a \textit{chromosome}. The collection of these $N$ is called a \textit{population}. In simple words, we initialize a population of $N$ chromosomes. Let $\{\mathbf{c}_1,...,\mathbf{c}_N\}$ be the initial population where $\mathbf{c}_i$ is the $i$th chromosome. A \textit{gene} is defined as a specific entry in the chromosome vector.

We next calculate the power values associated with these $N$ chromosomes using the power function value oracle. The power value of a chromosome is called its \textit{fitness}. Let $f_i$ be the fitness of $\mathbf{c}_i$. We save these chromosomes and the corresponding fitness values in a hash map (dictionary).

We next go to \textit{reproduction} to form the next generation of chromosomes. The chance of \textit{selection} of a chromosome from a past generation to the next generation is an increasing function of its fitness. One such selection policy is the $\lambda$-logistic policy where the probability of selecting a chromosome $u_i$ is calculated as follows.
\begin{align}
    u_i = \frac{\exp(\lambda f_i)}{\exp\left(\lambda\sum_{i=1}^{N}f_i\right)}
\end{align}

$\lambda$ is a hyper-parameter. 

The idea of using power value as fitness for reproduction is motivated by the fact that we are interested in maximizing the power and would like to travel/take steps towards the high-power region. 

We next go to \textit{mutation} where we the probability of a chromosome going through mutation is $p_m$ (a hyper-parameter). If a chromosome is selected for mutation then we mutate a randomly chosen gene of that chromosome by replacing its current value with some randomly chosen value within its initial range.

We next go to \textit{crossover} where we select the best two chromosomes in terms of fitness. We then select a random index and split both chromosomes at that index. Finally, we merge the front piece of the first chromosome with the end piece of the second chromosome, and vice-versa. 

We then repeat reproduction, mutation, and crossover for some $I$ (a hyper-parameter) iterations. Note that, in these successive iterations some chromosomes are repeated. We do not need to query the power value oracle again to calculate their fitness as we save this information in a dictionary. We update this dictionary after every iteration for new (not seen previously) chromosomes.

The \textit{final dictionary} with all \textit{(chromosome, power value)} pairs that the genetic algorithm comes across through the iterations of the genetic algorithm in the process of learning to reach a high/max power region gives us an \textit{estimate of the power manifold} of interest.

\begin{algorithm}
\caption{GeneticPower}\label{alg:geneticpower}
\begin{algorithmic}[1]
    \State \textbf{Input} \{$\boldsymbol{\theta}_l,\boldsymbol{\theta}_u, n_l, n_u, \phi, \alpha, \lambda, p_m, I$\}.
    \State Initialize a population $\{\mathbf{c}_1,...,\mathbf{c}_N\}$.
    \State Define $\mathcal{D} = \{\}$, empty dictionary.
    \State iter = 0
    \State \textbf{while} {iter $\le I$} \textbf{do}
        \State \qquad \textbf{for} {each $\mathbf{c}$ in $\{\mathbf{c}_1,...,\mathbf{c}_N\}$} \textbf{do}
            \State \qquad \qquad \textbf{if} {$\mathbf{c}$ in $\mathcal{D}$}
                \State \qquad \qquad \qquad Look up fitness($\mathbf{c}$) from $\mathcal{D}$.
            \State \qquad \qquad \textbf{else}
                \State \qquad \qquad \qquad fitness($\mathbf{c}$) = $f_{\phi,\alpha}(\mathbf{c})$ using power function value oracle.
            \State \qquad \qquad \textbf{end}
        \State \qquad \textbf{end}
        \State \qquad Append $\{(\mathbf{c}_i,f_i),i=1,...,N\}$ to $\mathcal{D}$.
        \State \qquad Reproduce using a $\lambda$-logistic policy.
        \State \qquad Mutate each chromosome w.p. $p_m$.
        \State \qquad Perform crossover.
        \State \qquad iter = iter + 1
    \State \textbf{end}
    \State \textbf{return} $\mathcal{D}$.
\end{algorithmic}
\end{algorithm}

\subsection{Power Prediction using Nearest Neighbors}

So far, we have estimated the power manifold using the genetic algorithm discussed in Algorithm \ref{alg:geneticpower}. We next answer the following question. How do we predict the power value for a new set of arguments? Note that the genetic algorithm described in Algorithm \ref{alg:geneticpower} comes across some sets of arguments but not all. It is indeed desirable that Algorithm \ref{alg:geneticpower} queries the power value oracle as less as possible but still be able to learn the manifold well. However, in general, a user may be interested in knowing the power function value for an arbitrary set of arguments. Once we have estimated/learned the power manifold, we can use this estimated manifold to predict the power values for an arbitrary set of arguments instead of querying the costly power value oracle. We use a simple nearest neighbors predictor described as follows.

For a given set of arguments $\mathbf{c}$, find $k$ (a hyper-parameter) nearest neighbors (in terms of Euclidean distance) in $\mathcal{D}$ returned by Algorithm \ref{alg:geneticpower}. Provide a prediction of the power for $\mathbf{c}$ as the average of the power values of those $k$ nearest neighbors in $\mathcal{D}$. The process is outlined in Algorithm \ref{alg:knn}.

\begin{algorithm}
\caption{kNearestNeighbors}\label{alg:knn}
\begin{algorithmic}[1]
    \State \textbf{Input} \{$\mathcal{D},\mathbf{c},k$\}.
    \State Find $k$ nearest neighbors of $\mathbf{c}$ in $\mathcal{D}$.
    \State Let $\mathbf{c}_{(1)},...,\mathbf{c}_{(k)}$ be those neighbors.
    \State Let $f_{(1)},...,f_{(k)}$ be the corresponding power values.
    \State \textbf{return} $\frac{1}{n}\sum_{i=1}^kf_{(i)}$.
\end{algorithmic}
\end{algorithm}

\begin{figure}
    \centering
    {\input{figures/methodology-figure.tex}}
    \caption{The proposed methodology with the genetic algorithm on the left and the nearest neighbors on the right.}
    \label{fig:system_model}
\end{figure}
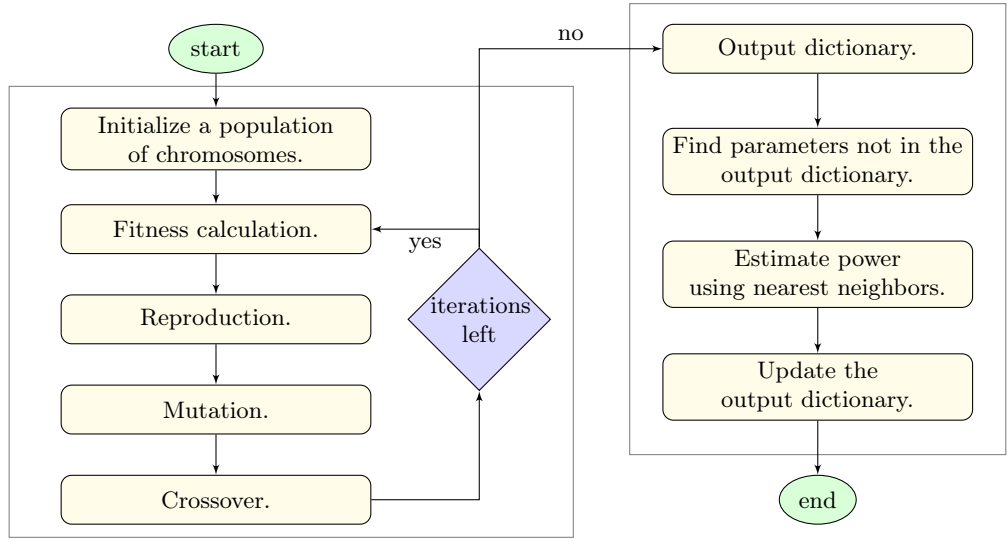

Refer to Figure \ref{fig:system_model} for an overview of the proposed methodology.

\subsection{Power Function Value Oracle}

In Algorithm \ref{alg:geneticpower}, we calculate the fitness of a chromosome as its power using a power function value oracle. We next discuss how can we create such an oracle.

Recall the definition of statistical power of a test as follows.
\begin{align*}
    1 - \gamma &= \text{Prob}(\text{Reject } H_0 \text{ when it is false.})
\end{align*}

Let $T(y_1,...,y_n)$ be the test statistic and $T_\alpha$ is the critical value at $\alpha$ level of significance associated with the hypothesis test under consideration. Statistical power can be calculated as follows.
\begin{align}
    1 - \gamma &= \text{Prob}(T(y_1,...,y_n) > T_\alpha | H_1 \text{ is true.})
\end{align}

Hence, to calculate the power of a test, we need the distribution of test statistic when the alternative hypothesis is true. Obtaining the non-null (alternative) distribution of the test statistic is not always easy. 

However, for a given set of values of the parameters $\boldsymbol{\theta}$ given the values of the parameters, power can be estimated using the following simulation-based approach. For some given (non-zero) values of the parameters, simulate the sample observations using these values and probability distribution $F$. For instance, in the case of multiple linear regression, it can be done by using the model equation. This generated data is from the model when the null hypothesis of no difference is false. Using this data, estimate the model parameters and test the hypothesis of interest. If the null hypothesis is rejected when it is false then these are the events that count towards the power of the test. Perform this simulation a large number of times then the power can be estimated as the relative frequency of rejections of the null hypothesis. The process is outlined in Algorithm \ref{alg:pfvo}.

\begin{algorithm}
\caption{PowerOracle}\label{alg:pfvo}
\begin{algorithmic}[1]
    \State \textbf{Input} \{$\boldsymbol{c}, F, \alpha,\text{nsim}$\}.
    \State Rejections = 0
    \State \textbf{for} {$i = 1,...,\text{nsim}$} \textbf{do}
        \State \qquad Generate sample observations using $\boldsymbol{c}$ and $F$.
        \State \qquad Test the hypothesis of interest.
        \State \qquad \textbf{if} {The test rejects the null hypothesis.}
            \State \qquad \qquad Rejections = Rejections + 1.
        \State \qquad \textbf{end}
    \State \textbf{end}
    \State \textbf{return} Rejections/nsim.
\end{algorithmic}
\end{algorithm}

\section{Applying {GeneticPower} to Multiple Linear Regression Model} 

A multiple linear regression model represents a relationship between a dependent $(y)$ and several independent variables $(x_1,...,x_p)$ as follows.
\begin{align}
    y_i = \beta_0 + \beta_1 x_{i1} + ... +\beta_p x_{ip} + \epsilon_{i}, \ i = 1,...,n, \\
    \epsilon_i \sim N(0,\sigma^2), \text{independently.} \notag
\end{align}

The coefficients $\beta_1,...,\beta_p$ are called the model parameters, and $\sigma^2$ is called the noise parameter. For simplicity, assume $\beta_0 = 0$.

Define, $\mathbf{y} = (y_1,....,y_n)^T$, $\mathbf{X} = [x_{ij}]_{i=1,...,n; j=1,...,p}$, $\boldsymbol\beta = (\beta_1,...,\beta_p)$, and $\boldsymbol\epsilon = (\epsilon_1,...,\epsilon_n)^T$.

Hence, a multiple linear regression model can be represented as follows.
\begin{align}
    \mathbf{y} = \mathbf{X}\boldsymbol{\beta} + \boldsymbol{\epsilon}, \ \boldsymbol{\epsilon} \sim \mathbf{N}_n(\mathbf{0},\sigma^2\mathbf{I}) 
\end{align}
We estimate $\boldsymbol\beta$ using the least squares method. Further, we test for the statistical significance of individual regressors and the overall regression. In fact, we can test all different kinds of hypotheses involving model parameters. 

\textit{Testing for the significance of individual regressors.} The null and alternative hypotheses of interest are as follows. $H_0: \beta_i = 0$ against $H_1: \beta_i \ne 0$. We use a $t$-test to test the significance of the partial regression coefficient $\beta_i$ while controlling for the rest of the model parameters.

\textit{Testing for the joint significance of some regressors.} The null and alternative hypotheses of interest are as follows. $H_0: \beta_{i_1} =...= \beta_{i_k} = 0$ against $H_1: \text{At least one of } \beta_{i_1}, ..., \beta_{i_k} \text{ is different from zero}$, where $\{i_1,...,i_k\}$ is some subset of $\{1,...,p\}$. We use an $F$-test to test the significance of the joint significance of $\beta_{i_1},...,\beta_{i_k}$ while controlling for the rest of the model parameters.

\textit{Testing for the significance of overall regression.} The null and alternative hypotheses of interest are as follows. $H_0: \beta_{1} =...= \beta_{p} = 0$ against $H_1: \text{At least one of } \beta_{1}, ..., \beta_{p} \text{ is different from zero}$. We use an $F$-test to test the significance of the joint significance of $\beta_{1},...,\beta_{p}$.

Note that testing for the significance of overall regression is a special case of testing for the joint significance of some regressors when $\{i_1,...,i_k\} = \{1,...,p\}$.

\begin{algorithm}
\caption{MLR-PowerOracle}\label{alg:pfvo-mlr}
\begin{algorithmic}[1]
    \State \textbf{Input} \{$\boldsymbol\beta,\alpha,n,\sigma^2,\text{nsim}$\}.
    \State Rejections = 0
    \State \textbf{for} $i = 1,...,\text{nsim}$ \textbf{do}
        \State \qquad Generate $\boldsymbol{\epsilon}$ from $\mathbf{N}_n(\mathbf{0},\sigma^2\mathbf{I})$.
        \State \qquad Generate $\mathbf{x}_1,..,\mathbf{x}_n$ from $\mathbf{N}_n(\mathbf{0},\mathbf{I})$.
        \State \qquad Create $\mathbf{X} = [\mathbf{x}_1,..,\mathbf{x}_n]$.
        \State \qquad Calculate $\mathbf{y} = \mathbf{X}\boldsymbol{\beta} + \boldsymbol{\epsilon}$.
        \State \qquad Fit a regression model using $\mathbf{y}$ and $\mathbf{X}$.
        \State \qquad Test the hypothesis of interest.
        \State \qquad \textbf{if} {The test rejects the null hypothesis.}
            \State \qquad \qquad Rejections = Rejections + 1
        \State \qquad \textbf{end}
    \State \textbf{end}
    \State \textbf{return} Rejections/nsim
\end{algorithmic}
\end{algorithm}

For a multiple linear regression model, an arbitrary chromosome is given as $\mathbf{c} = (\beta_1,...,\beta_p,n)$ for the above three tests. In the context of a multiple linear regression model, the power function value oracle is outlined in Algorithm \ref{alg:pfvo-mlr}. After learning the power manifold using Algorithm \ref{alg:geneticpower}, we can do the nearest neighbors prediction using Algorithm \ref{alg:knn}

%% file: figures/methodology-figure.tex
    \begin{tikzpicture}[node distance = 1.2cm, auto]
        \node[rectangle,
        draw = gray,
        text = olive,
        minimum width = 7.5cm, 
        minimum height = 6cm] (r) at (1,-3.5) {};
        \node[rectangle,
        draw = gray,
        text = olive,
        minimum width = 5cm, 
        minimum height = 6cm] (r) at (8,-2.4) {};
        \node [cloud] (begin) {start};
        \node [block, below of=begin] (init) {Initialize a population \\ of chromosomes.};
        \node [block, below of=init] (fitness) {Fitness calculation.};
        \node [block, below of=fitness] (reproduce) {Reproduction.};
        \node [block, below of=reproduce] (mutation) {Mutation.};
        \node [block, below of=mutation] (crossover) {Crossover.};
        \node [decision, right of=reproduce] (decide) {iterations left};
        \node [block, right of=begin, node distance = 8cm] (output) {Output dictionary.};
        \node [block, below of=output, node distance=1.5cm] (find) {Find parameters not in the \\ output dictionary.};
        \node [block, below of=find, node distance=1.5cm] (knn) {Estimate power \\ using nearest neighbors.};
        \node [block, below of=knn, node distance=1.5cm] (update) {Update the \\ output dictionary.};
        \node [cloud, below of=knn, node distance = 3cm] (end) {end};
        \path [line] (begin) -- (init);
        \path [line] (init) -- (fitness);
        \path [line] (fitness) -- (reproduce);
        \path [line] (reproduce) -- (mutation);
        \path [line] (mutation) -- (crossover);
        \path [line] (decide) |- node [near end] {no} (output);
        \path [line] (crossover) -| (decide);
        \path [line] (decide) |- node [near end] {yes} (fitness);
        \path [line] (output) -- (find);
        \path [line] (find) -- (knn);
        \path [line] (knn) -- (update);
        \path [line] (update) -- (end);
    \end{tikzpicture}

%% file: sections/experiments.tex
\section{Experiments} \label{sec:experiments}

In this section, we provide experiments to demonstrate the performance of our framework in terms of power function estimation and run-time. We consider the following multiple linear regression model.
\begin{align}
    y = \beta_0 + \beta_1x_1 + \beta_2x_2 + \beta_3(x_1x_2) + \epsilon,
\end{align}

where $y$ is the response, $x_1$ is the experimental condition (-1: control, 1: treatment), $x_2$ is other measure, and  $x_1x_2$ is the interaction of the experimental condition and the other measure. 

Based on similar prior knowledge we know the following. Only experimental condition $x_1$ main effects of at least 0.10 would be substantively meaningful, but expect that the population parameter could be as large as 0.30. The main effect of the other measure $x_2$ could range from 0.30 to 0.90. We don’t have a very good sense of interaction magnitudes, although it would need to be greater than  zero. We only have the budget to afford 500 individuals at the maximum. The level of significance is 0.05. This translates to $\beta_1 \in [0.10, 0.30]$, $\beta_2 \in [0.30, 0.90]$, $\beta_3 > 0$, $n \le 500$, $\alpha = 0.05$.

For GeneticPower, we use, $N=1000, 2000, ..., 5000$, $I=10, 20, ..., 100$, $\lambda = 1$, $p_m = 0.05$, regression coefficients' discretization step size = 0.05, and ample size's discretization step size = 5. For MLR-PowerOracle, we use, nsim = 1000. For kNearestNeighbors, we use  $k=5$. 

For our experiments, we focus on two tests, viz. (i) $H_0: \beta_1 = 0$ against $H_1: \beta_1 \ne 0$, and (ii) $H_0: \beta_3 = 0$ against $H_1: \beta_3 \ne 0$. We use a $t$-test to test the significance $\beta_i, i = 1,3$ while controlling for the rest of the model parameters.

For assessing the performance of our algorithm, we compute the power manifold using a \textit{brute-force} method which involves dividing the initial ranges of the regression coefficients into a high-dimensional grid and computing the power using the power function value oracle discussed in Algorithm \ref{alg:pfvo-mlr}. For of creating the grid, we use the discretization step size for regression coefficients as 0.05 and sample size as 5.

We use GeneticPower (Algorithm \ref{alg:geneticpower}) to learn/estimate the power manifold. For the set of arguments that are seen by the brute-force method but not by the genetic algorithm, we predict the power $k$-nearest neighbors (Algorithm \ref{alg:knn}). The experiments are carried out on a computer with 2.6 GHz 24-core Intel Xeon Gold Sky Lake processor and 96 GB of memory. We used Python for our implementation. The source code for our experiments is available \href{https://anonymous.4open.science/r/geneticpower-7C6D/}{here}. 

We compare the performance of our algorithm with a costly brute-force strategy in terms of the quality of the estimation using root mean squared error. Let $\mathbf{c}_1,...,\mathbf{c}_M$ be all set of arguments seen by the brute-force method. Let $f^{(b)}_1,....f^{(b)}_M$ be the corresponding power values computed using the brute-force method. Let $f^{(g)}_1,....f^{(g)}_M$ be the corresponding power values using our algorithm (either after the genetic algorithm or after the nearest neighbors prediction). We calculate the mean squared error (RMSE) as follows.
\begin{align}
    \text{RMSE} = \left[\frac{1}{n}\sum_{i=1}^M\left(f^{(b)}_i-f^{(g)}_i\right)^2\right]^{1/2}
\end{align}

For our experiments, computing power manifold for $H_0: \beta_1 = 0$ against $H_1: \beta_1 \ne 0$, and $H_0: \beta_3 = 0$ against $H_1: \beta_3 \ne 0$ using the brute-force method takes approximately 8000 seconds. 

For different population sizes, the time taken by our algorithm and root mean squared error as a function of the number of iterations are plotted in Figure \ref{fig:runtimeandrmse}. We make the following observations. The time taken by our algorithm is always less than the brute-force method. The time taken by our algorithm increases as the number of iterations increases and the size of the population increases, respectively. The RMSE for our algorithm decreases as the number of iterations increases and the size of the population increases, respectively. The rate of increase in run-time as a function of the number of iterations decreases as the size of the population increases. The rate of decrease in RMSE as a function of the number of iterations decreases as the size of the population increases. The reason for these behaviors is the following. A smaller (larger) initial population will require less (more) time in calculating the fitness of the initial population but in the future the algorithm will come across more (less) new set of arguments.

For different population sizes, the shape of the learned power manifold for $H_0: \beta_3 = 0$ against $H_1:    \beta_3 \ne 0$ as a function of the number of iterations are provided in Figures \ref{fig:qualityoflearning31000}-\ref{fig:qualityoflearning35000}. In all the sub-figures, the effect size ($\beta_3$), sample size ($n$), and power ($1-\gamma$) are plotted on $x$-axis, $y$-axis and $z$-axis, respectively. We make the following observations.
The quality of learning in terms of closeness to the brute-force manifold (shown in Figure \ref{fig:bruteforcepowermanifold}) improves as the number of iterations increases and the size of the population increases. Moreover, the proposed algorithm has been able to learn the shape of the manifold fairly quickly (in terms of run time). The improvement in the quality of learning is increasing as the population size increases. This is the same as saying that the rate of decrease in RMSE as a function of the number of iterations decreases as the size of the population increases from Figure \ref{fig:runtimeandrmse}. The quality of learning the power manifold is better for the set of arguments with high values of power as compared to the ones with low values of power. This is because the selection scheme during the reproduction step favors the high-power set of arguments more as compared to the low-power ones.

For different population sizes, for learning the power manifold for $H_0: \beta_1 = 0$ against $H_1: \beta_1 \ne 0$, we obtain insights similar to the earlier case. We skipped the plots for this case for brevity. The actual power manifold in this case is not very complex and hence the proposed algorithm was able to learn it very quickly. However, for a highly complex manifold as in the earlier case as well the algorithm does learn the shape of the manifold fairly quickly.

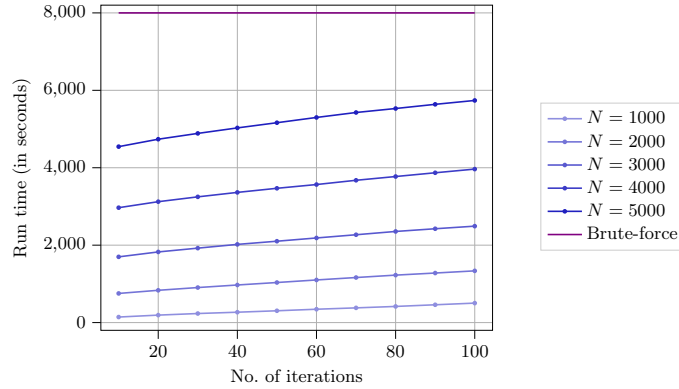
\begin{figure*}[t]
\centering
  \resizebox{.75\textwidth}{!}{\input{figures/run_times_all}}
\caption{Run times vs. no. of iteration. \textit{Brute-force run-time is 8000 seconds.}} \label{fig:runtime}
\end{figure*}

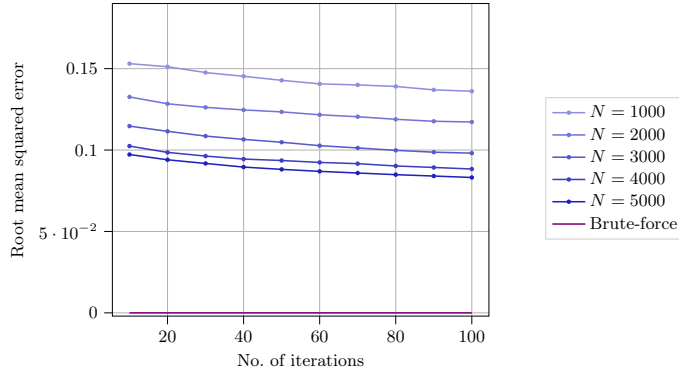
\begin{figure*}[t]
\centering
  \resizebox{.75\textwidth}{!}{\input{figures/rmses_all}}
\caption{Root mean squared error vs. no. of iteration. \textit{Brute-force RMSE is zero.}} \label{fig:rmse}
\end{figure*}

\input{sections/plots_beta3}

%% file: figures/run_times_all.tex
\begin{tikzpicture}
\colorlet{color0}{red!50!green!25!blue!100}
\colorlet{color1}{red!100!green!50!blue!}

\begin{axis}[
legend cell align={left},
legend style={
  fill opacity=0.8,
  draw opacity=1,
  text opacity=1,
  at={(1.5,0.7)},
  anchor=north east,
  draw=white!80!black
},
tick align=outside,
tick pos=left,
x grid style={white!69.0196078431373!black},
xlabel={No. of iterations},
xmajorgrids,
xmin=5.5, xmax=104.5,
xtick style={color=black},
y grid style={white!69.0196078431373!black},
ylabel={Run time (in seconds)},
ymajorgrids,
ymin=-200, ymax=8200,
ytick style={color=black}
]
\addplot [thick, color0!50, mark=*, mark size=.75, mark options={solid}]
table {%
10 143
20 193
30 234
40 268
50 305
60 345
70 380
80 417
90 460
100 503
};
\addlegendentry{$N=1000$}
\addplot [thick, color0!62.5, mark=*, mark size=.75, mark options={solid}]
table {%
10 753
20 834
30 905
40 971
50 1035
60 1101
70 1164
80 1226
90 1280
100 1336
};
\addlegendentry{$N=2000$}
\addplot [thick, color0!75, mark=*, mark size=.75, mark options={solid}]
table {%
10 1698
20 1825
30 1923
40 2020
50 2102
60 2186
70 2269
80 2354
90 2424
100 2491
};
\addlegendentry{$N=3000$}
\addplot [thick, color0!87.5, mark=*, mark size=.75, mark options={solid}]
table {%
10 2968
20 3123
30 3248
40 3363
50 3469
60 3566
70 3676
80 3773
90 3870
100 3965
};
\addlegendentry{$N=4000$}
\addplot [thick, color0!100, mark=*, mark size=.75, mark options={solid}]
table {%
10 4545
20 4737
30 4887
40 5029
50 5163
60 5299
70 5427
80 5530
90 5638
100 5738
};
\addlegendentry{$N=5000$}
\addplot [thick, color1, mark=*, mark size=0, mark options={solid}]
table {%
10 8000
20 8000
30 8000
40 8000
50 8000
60 8000
70 8000
80 8000
90 8000
100 8000
};
\addlegendentry{Brute-force}
\end{axis}

\end{tikzpicture}

%% file: figures/rmses_all.tex
\begin{tikzpicture}
\colorlet{color0}{red!50!green!25!blue!100}
\colorlet{color1}{red!100!green!50!blue!}

\begin{axis}[
legend cell align={left},
legend style={
  fill opacity=0.8,
  draw opacity=1,
  text opacity=1,
  at={(1.15,0.7)},
  anchor=north west,
  draw=white!80!black
},
tick align=outside,
tick pos=left,
x grid style={white!69.0196078431373!black},
xlabel={No. of iterations},
xmajorgrids,
xmin=5.5, xmax=104.5,
xtick style={color=black},
y grid style={white!69.0196078431373!black},
ylabel={Root mean squared error},
ymajorgrids,
ymin=-0.002, ymax=0.19,
ytick style={color=black},
]
\addplot [thick, color0!50, mark=*, mark size=.75, mark options={solid}]
table {%
10 0.153074383735657
20 0.151167273521423
30 0.147595763206482
40 0.145301699638367
50 0.142834424972534
60 0.140622735023499
70 0.139997363090515
80 0.139057636260986
90 0.136962413787842
100 0.136125564575195
};
\addlegendentry{ $N=1000$}
\addplot [thick, color0!62.5, mark=*, mark size=.75, mark options={solid}]
table {%
10 0.132611513137817
20 0.128418445587158
30 0.126214265823364
40 0.124638795852661
50 0.12343168258667
60 0.121659636497498
70 0.120468974113464
80 0.118874192237854
90 0.117678046226501
100 0.117221236228943
};
\addlegendentry{ $N=2000$}
\addplot [thick, color0!75, mark=*, mark size=.75, mark options={solid}]
table {%
10 0.114697694778442
20 0.111505389213562
30 0.108574390411377
40 0.106549978256226
50 0.104793787002563
60 0.102689027786255
70 0.101288318634033
80 0.0998198986053467
90 0.0986794233322144
100 0.0981060266494751
};
\addlegendentry{ $N=3000$}
\addplot [thick, color0!87.5, mark=*, mark size=.75, mark options={solid}]
table {%
10 0.102437973022461
20 0.0985563993453979
30 0.0962926149368286
40 0.0944653749465942
50 0.093572735786438
60 0.0923924446105957
70 0.0916244983673096
80 0.090222954750061
90 0.0893123149871826
100 0.0883481502532959
};
\addlegendentry{$N=4000$}
\addplot [thick, color0!100, mark=*, mark size=.75, mark options={solid}]
table {%
10 0.0972610712051392
20 0.0940103530883789
30 0.0917608737945557
40 0.0895441770553589
50 0.0881285667419434
60 0.0869154930114746
70 0.0858931541442871
80 0.0848793983459473
90 0.0840305089950562
100 0.0831540822982788
};
\addlegendentry{$N=5000$}
\addplot [thick, color1, mark=*, mark size=0, mark options={solid}]
table {%
10 0
20 0
30 0
40 0
50 0
60 0
70 0
80 0
90 0
100 0
};
\addlegendentry{Brute-force}
\end{axis}
\end{tikzpicture}

%% file: sections/plots_beta3.tex
\begin{figure*}[t]
\centering
\begin{minipage}{.65\textwidth}
  \centering
  \resizebox{\textwidth}{!}{\includegraphics{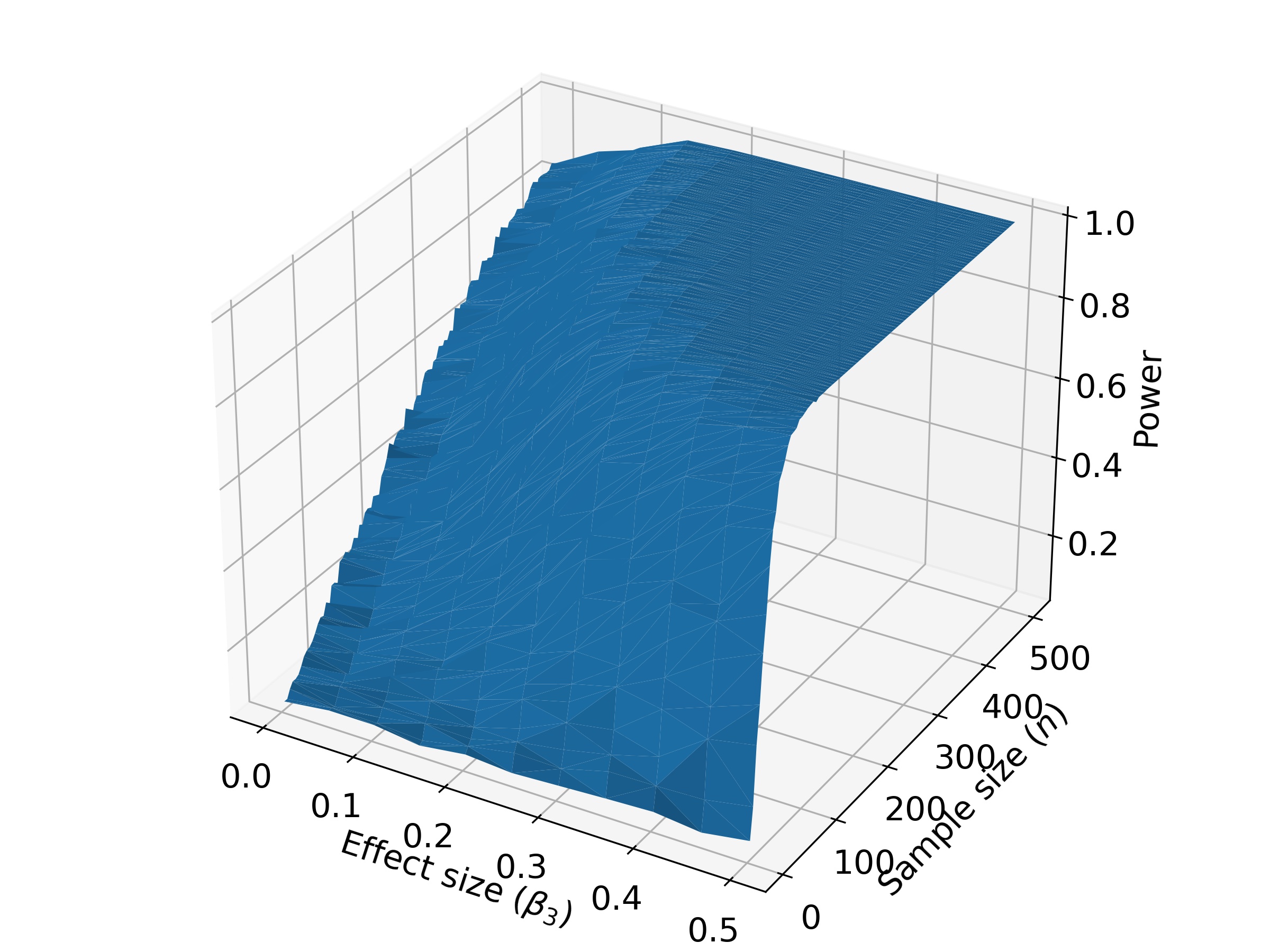}}
\end{minipage}
\caption{Brute-Force power manifold.}  \label{fig:bruteforcepowermanifold}
\end{figure*}

\begin{figure*}[t]
\centering
\begin{minipage}{.32\textwidth}
  \centering
  \resizebox{\textwidth}{!}{\includegraphics{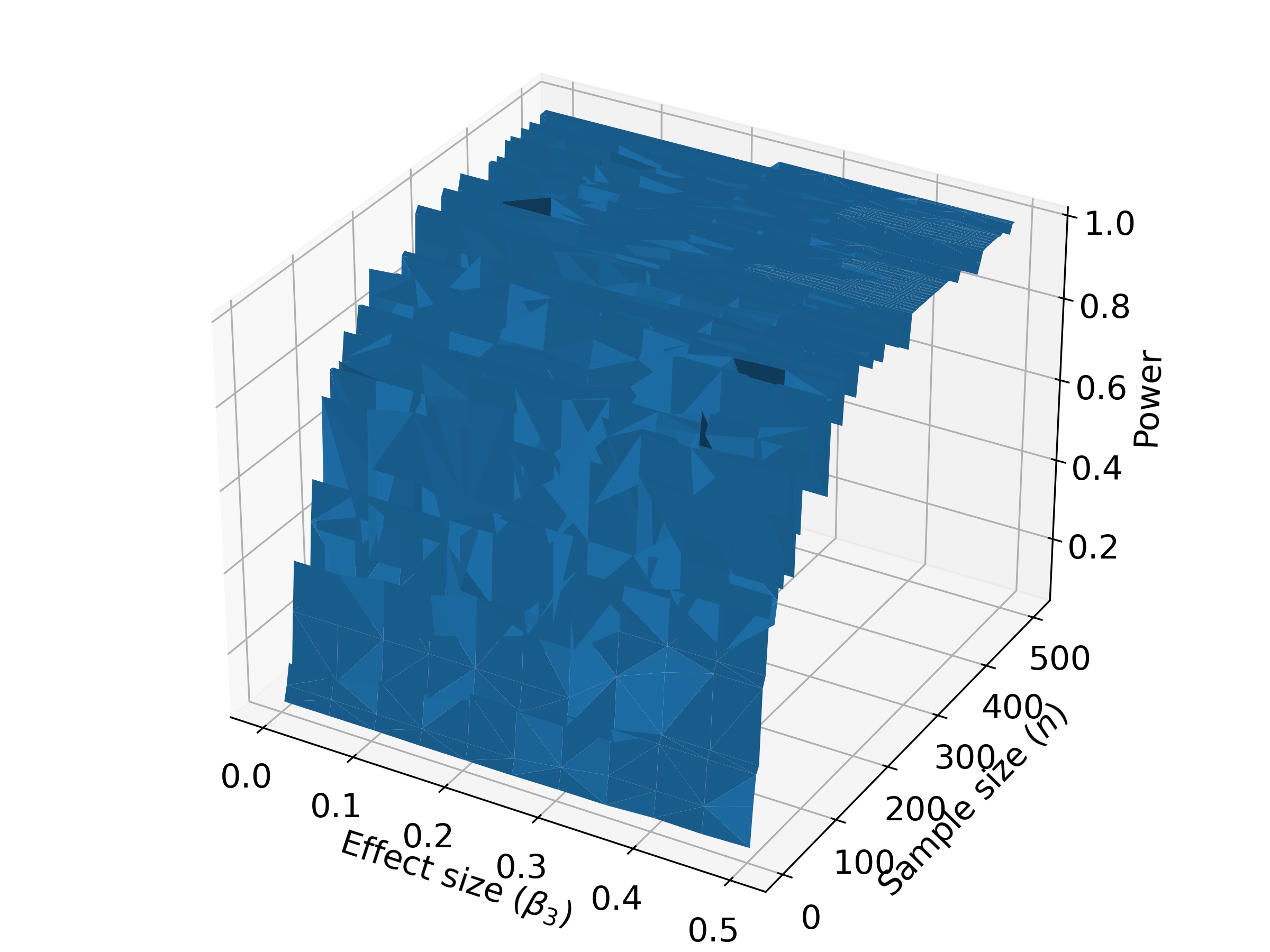}}
  \caption*{$I$ = 10}
\end{minipage}
\begin{minipage}{.32\textwidth}
  \centering
  \resizebox{\textwidth}{!}{\includegraphics{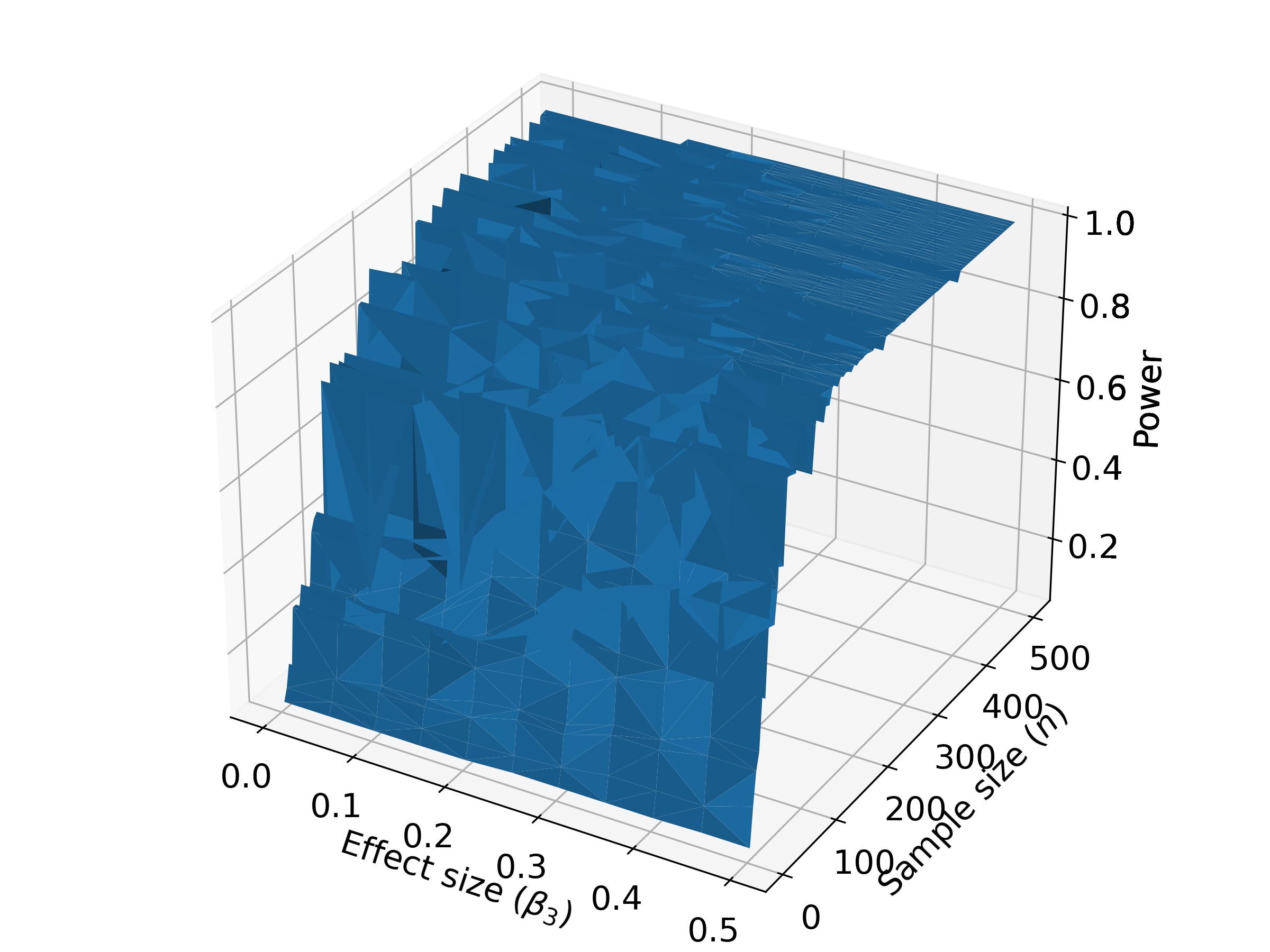}}
  \caption*{$I$ = 50}
\end{minipage}
\begin{minipage}{.33\textwidth}
  \centering
  \resizebox{\textwidth}{!}{\includegraphics{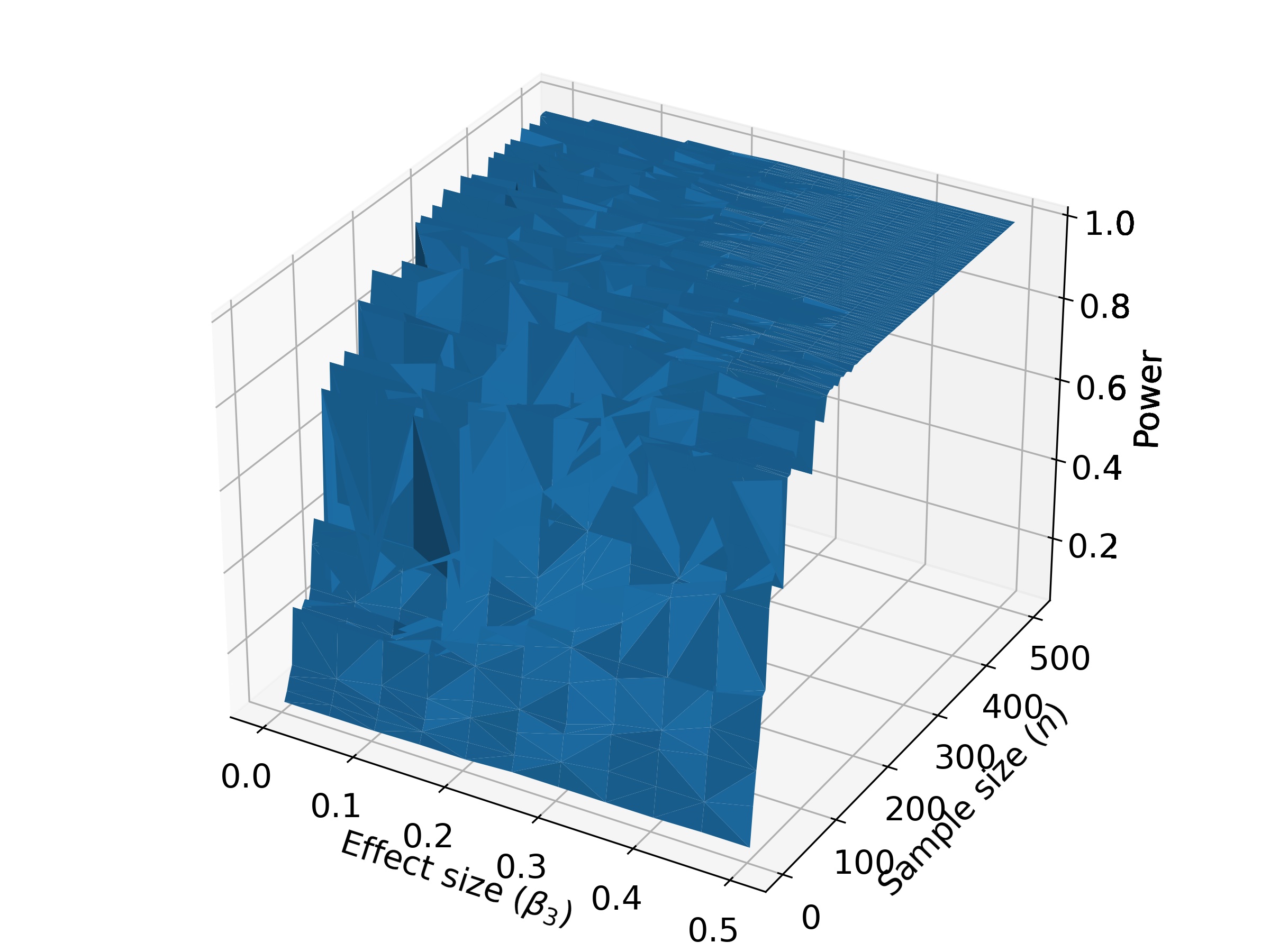}}
  \caption*{$I$ = 100}
\end{minipage}
\caption{Quality of learning the power manifold as a function of the effect size $\beta_3$ and sample size $n$ vs. the number of iterations $I$ when the population $N=1000$.} \label{fig:qualityoflearning31000}
\end{figure*}

\begin{figure*}[t]
\centering
\begin{minipage}{.32\textwidth}
  \centering
  \resizebox{\textwidth}{!}{\includegraphics{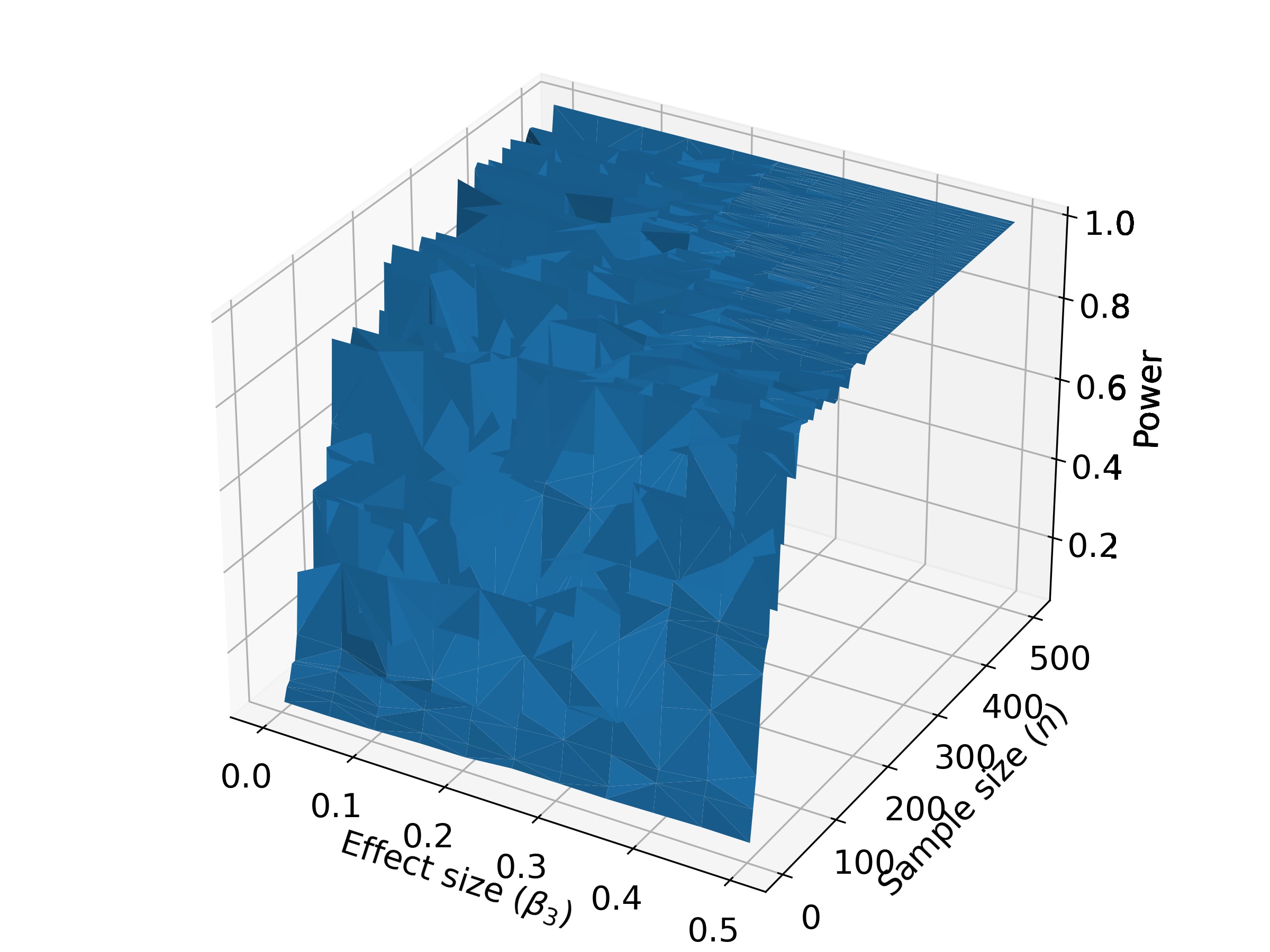}}
  \caption*{$I$ = 10}
\end{minipage}
\begin{minipage}{.32\textwidth}
  \centering
  \resizebox{\textwidth}{!}{\includegraphics{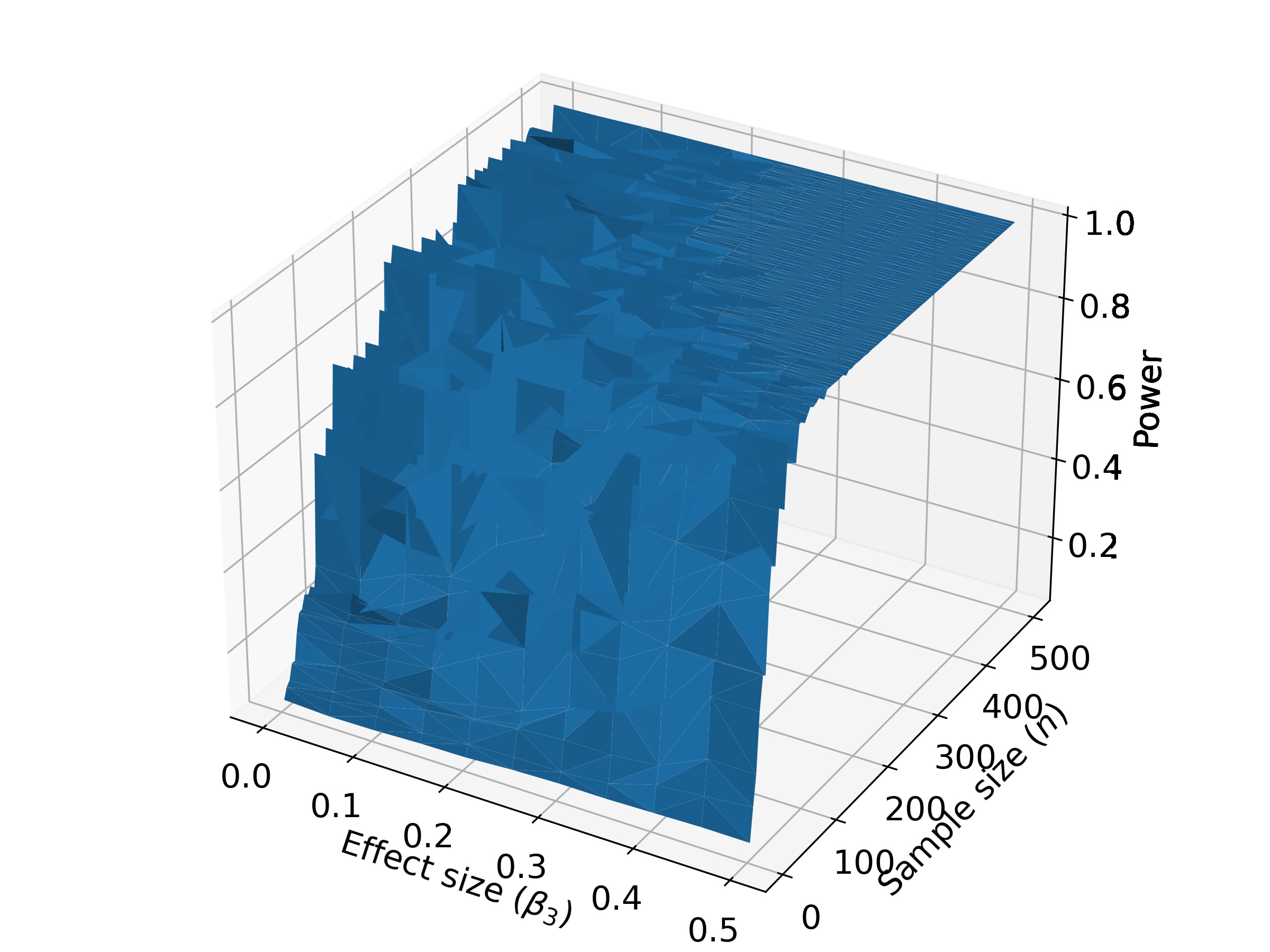}}
  \caption*{$I$ = 50}
\end{minipage}
\begin{minipage}{.33\textwidth}
  \centering
  \resizebox{\textwidth}{!}{\includegraphics{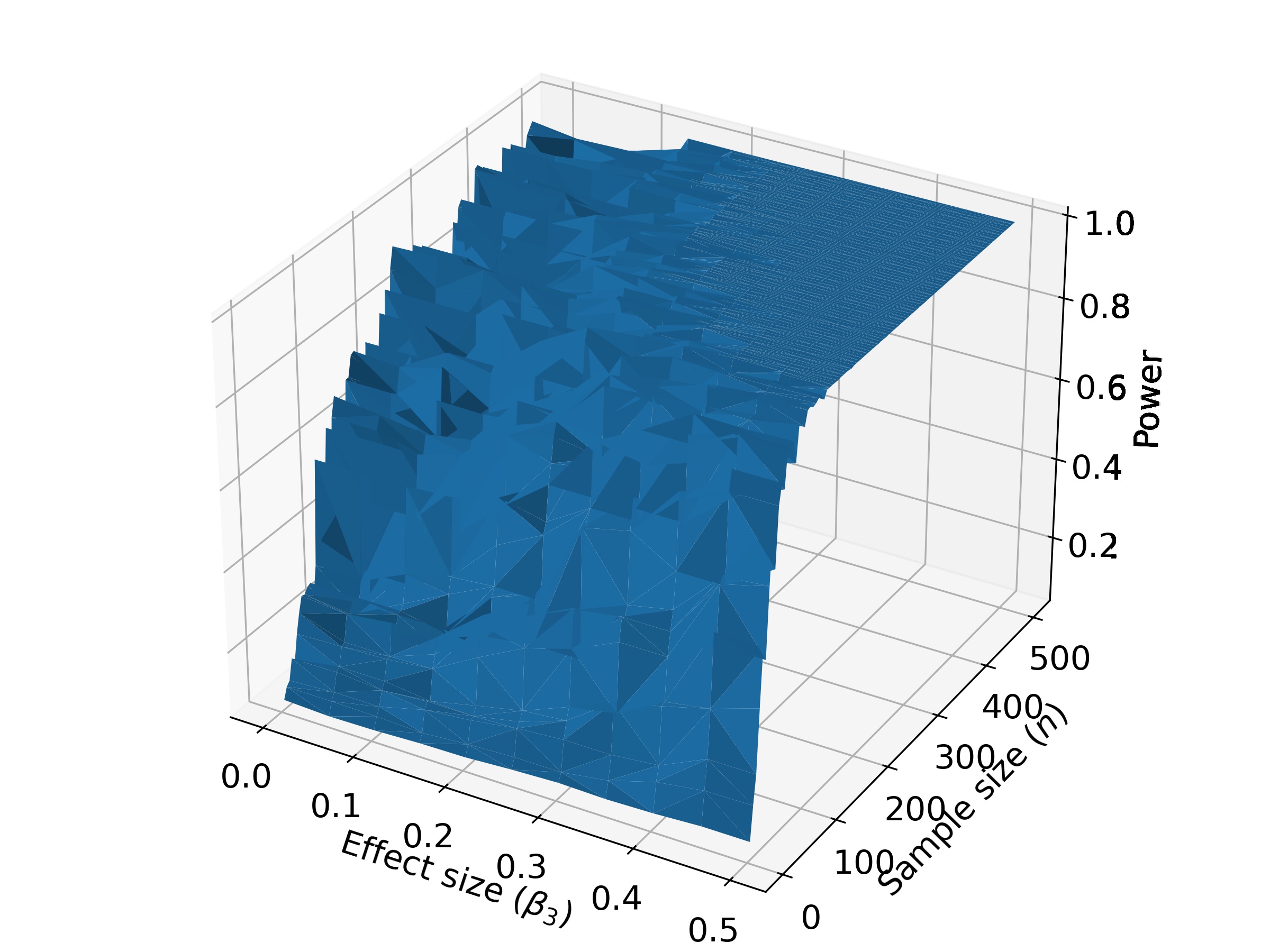}}
  \caption*{$I$ = 100}
\end{minipage}
\caption{Quality of learning the power manifold as a function of the effect size $\beta_3$ and sample size $n$ vs. the number of iterations $I$ when the population $N=2000$.}  \label{fig:qualityoflearning32000}
\end{figure*}

\begin{figure*}[t]
\centering
\begin{minipage}{.32\textwidth}
  \centering
  \resizebox{\textwidth}{!}{\includegraphics{figures/power_vs_beta3_and_sample_size_2000_10.jpg}}
  \caption*{$I$ = 10}
\end{minipage}
\begin{minipage}{.32\textwidth}
  \centering
  \resizebox{\textwidth}{!}{\includegraphics{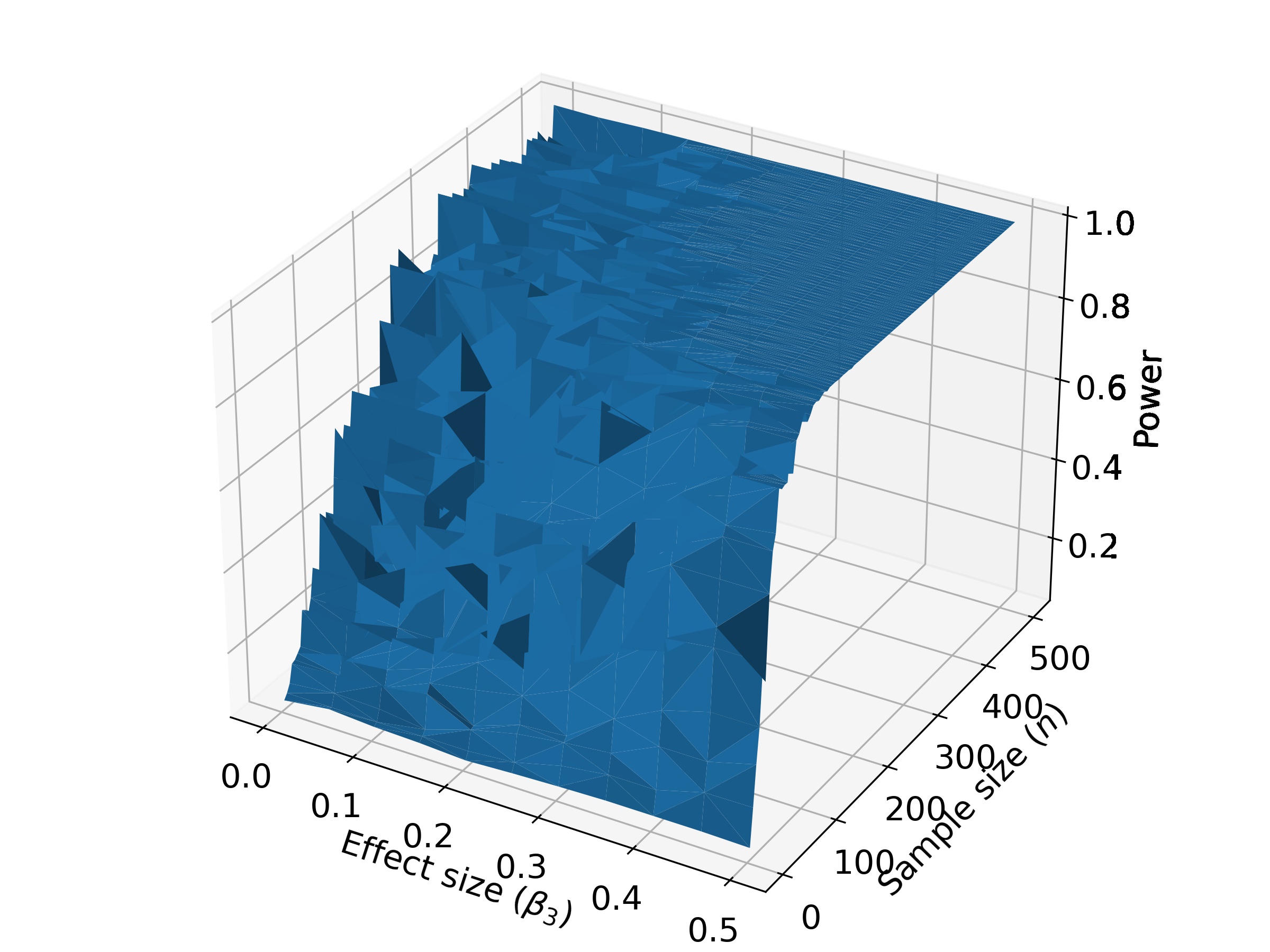}}
  \caption*{$I$ = 50}
\end{minipage}
\begin{minipage}{.33\textwidth}
  \centering
  \resizebox{\textwidth}{!}{\includegraphics{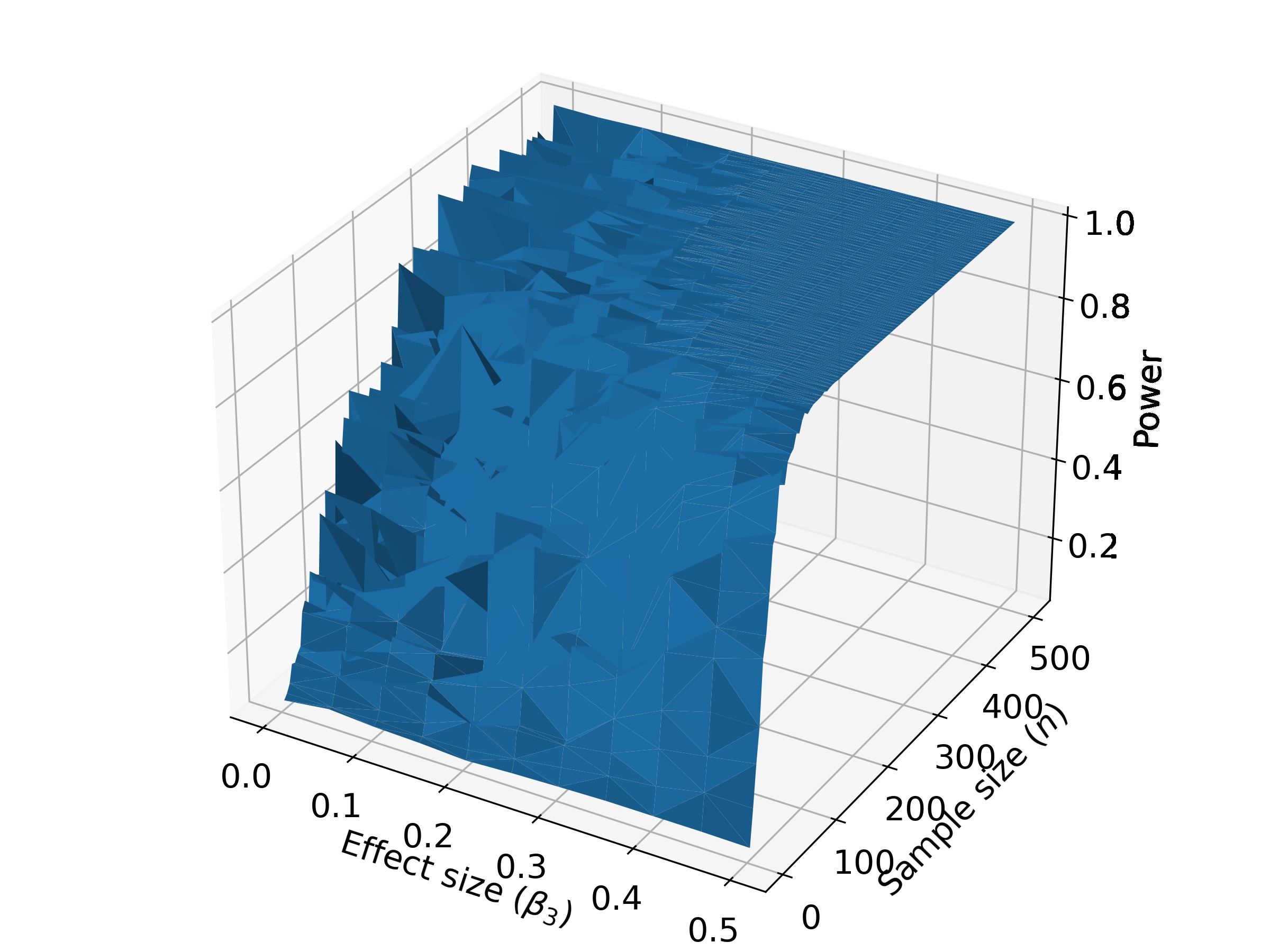}}
  \caption*{$I$ = 100}
\end{minipage}
\caption{Quality of learning the power manifold as a function of the effect size $\beta_3$ and sample size $n$ vs. the number of iterations $I$ when the population $N=3000$.}  \label{fig:qualityoflearning33000}
\end{figure*}

\begin{figure*}[t]
\centering
\begin{minipage}{.32\textwidth}
  \centering
  \caption*{$I$ = 10}
  \resizebox{\textwidth}{!}{\includegraphics{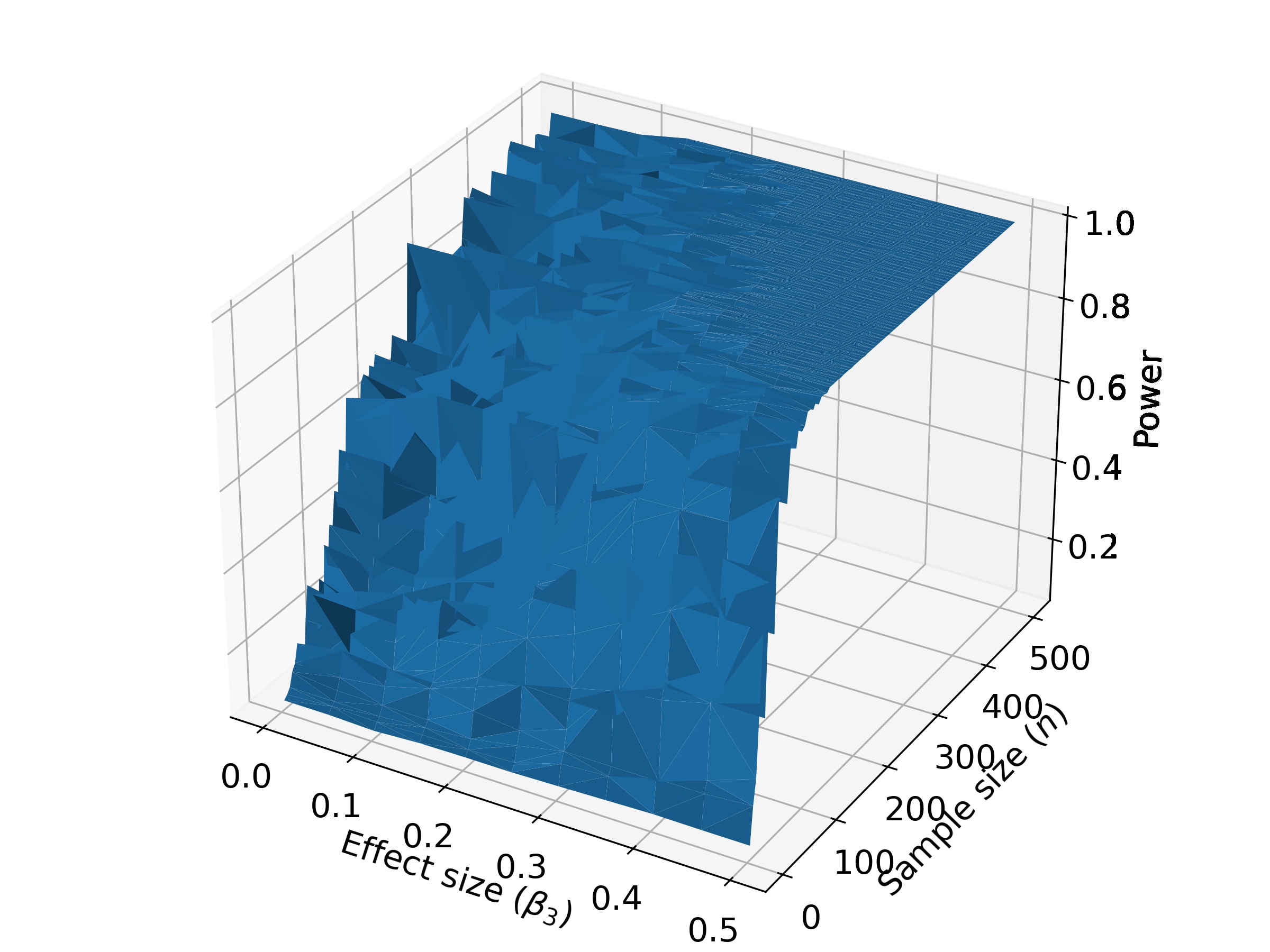}}
\end{minipage}
\begin{minipage}{.32\textwidth}
  \centering
  \caption*{$I$ = 50}
  \resizebox{\textwidth}{!}{\includegraphics{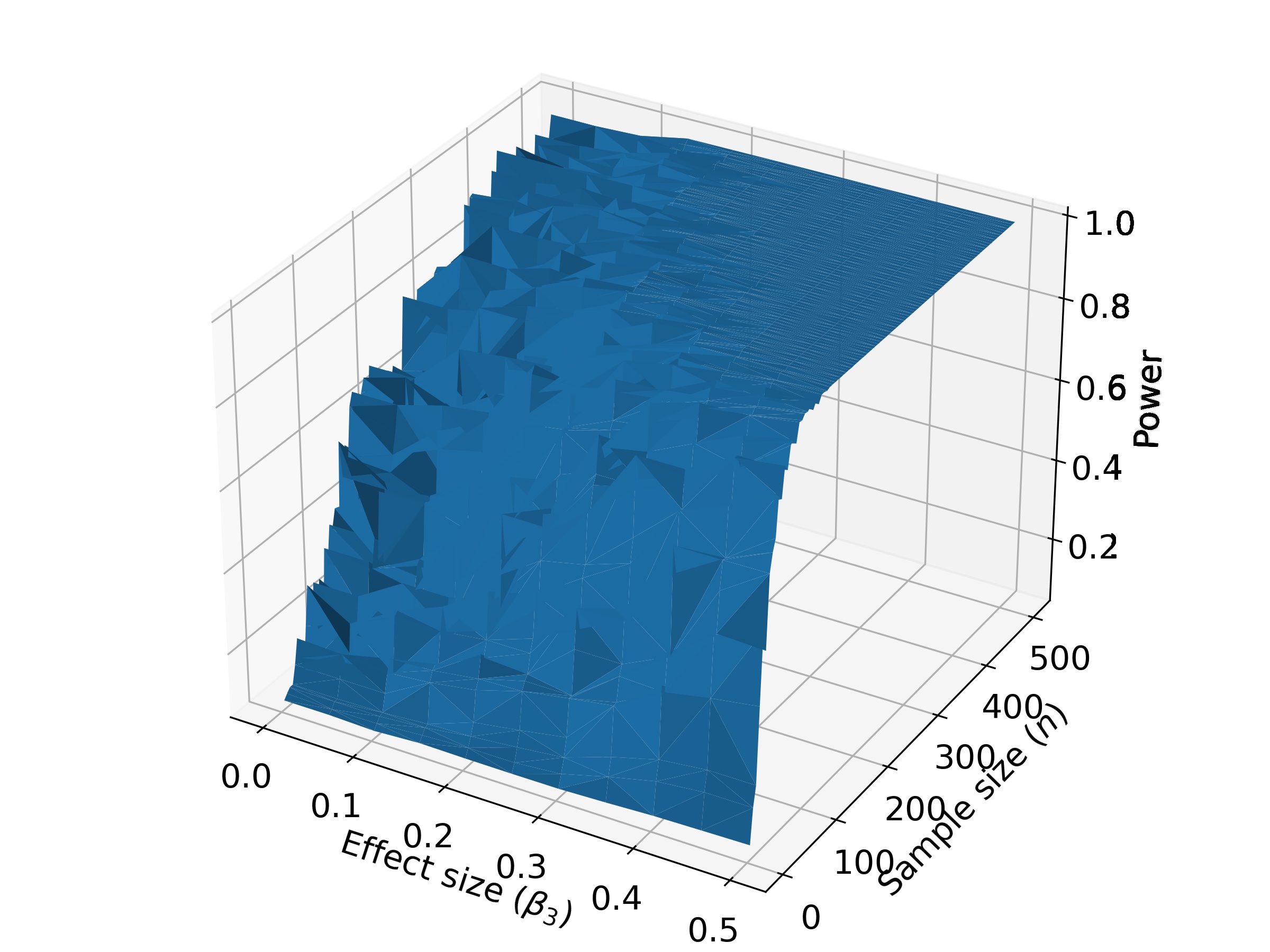}}
\end{minipage}
\begin{minipage}{.33\textwidth}
  \centering
  \caption*{$I$ = 100}
  \resizebox{\textwidth}{!}{\includegraphics{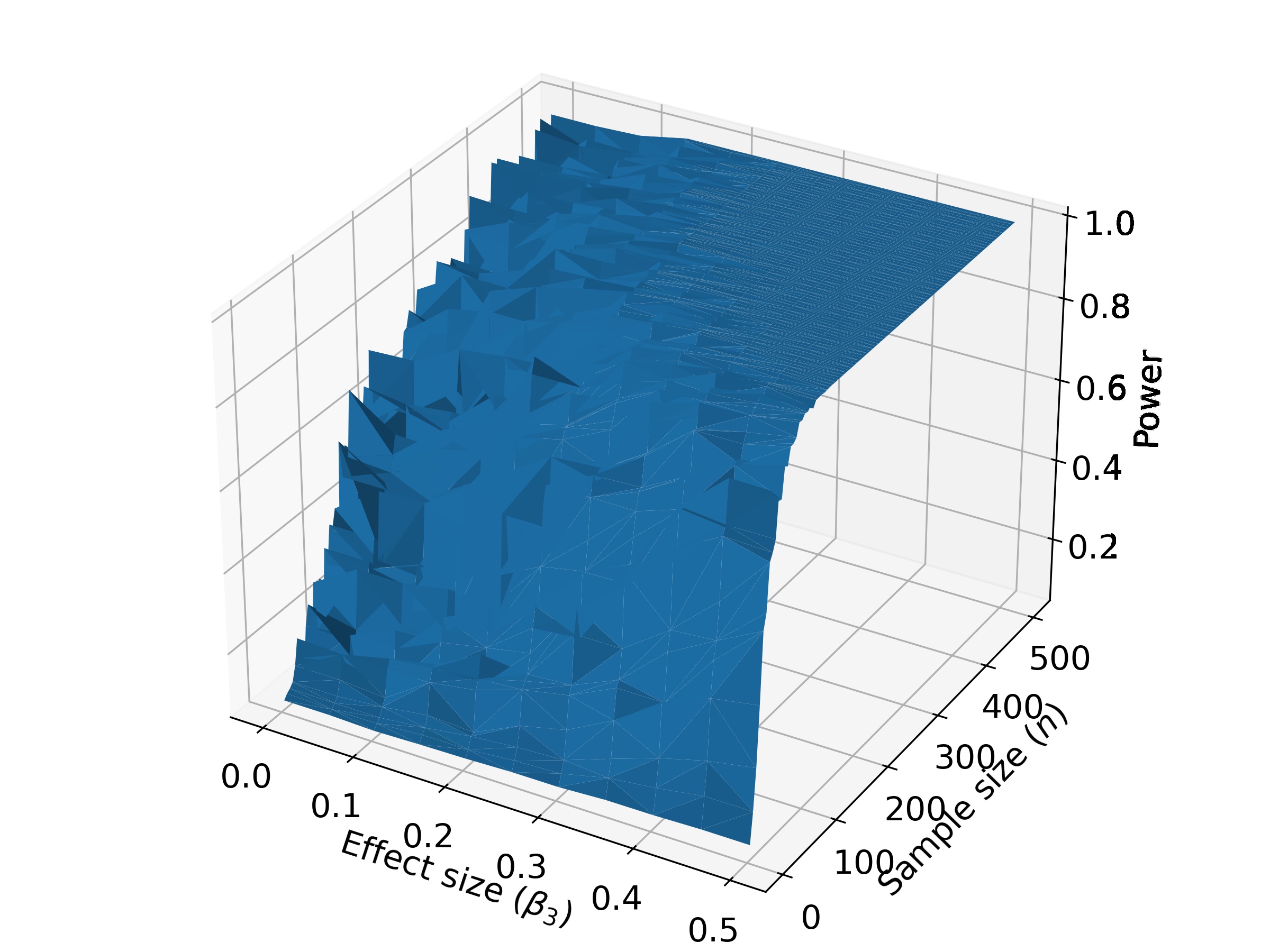}}
\end{minipage}
\caption{Quality of learning the power manifold as a function of the effect size $\beta_3$ and sample size $n$ vs. the number of iterations $I$ when the population $N=4000$.}  \label{fig:qualityoflearning34000}
\end{figure*}

\begin{figure*}[t]
\centering
\begin{minipage}{.32\textwidth}
  \centering
  \resizebox{\textwidth}{!}{\includegraphics{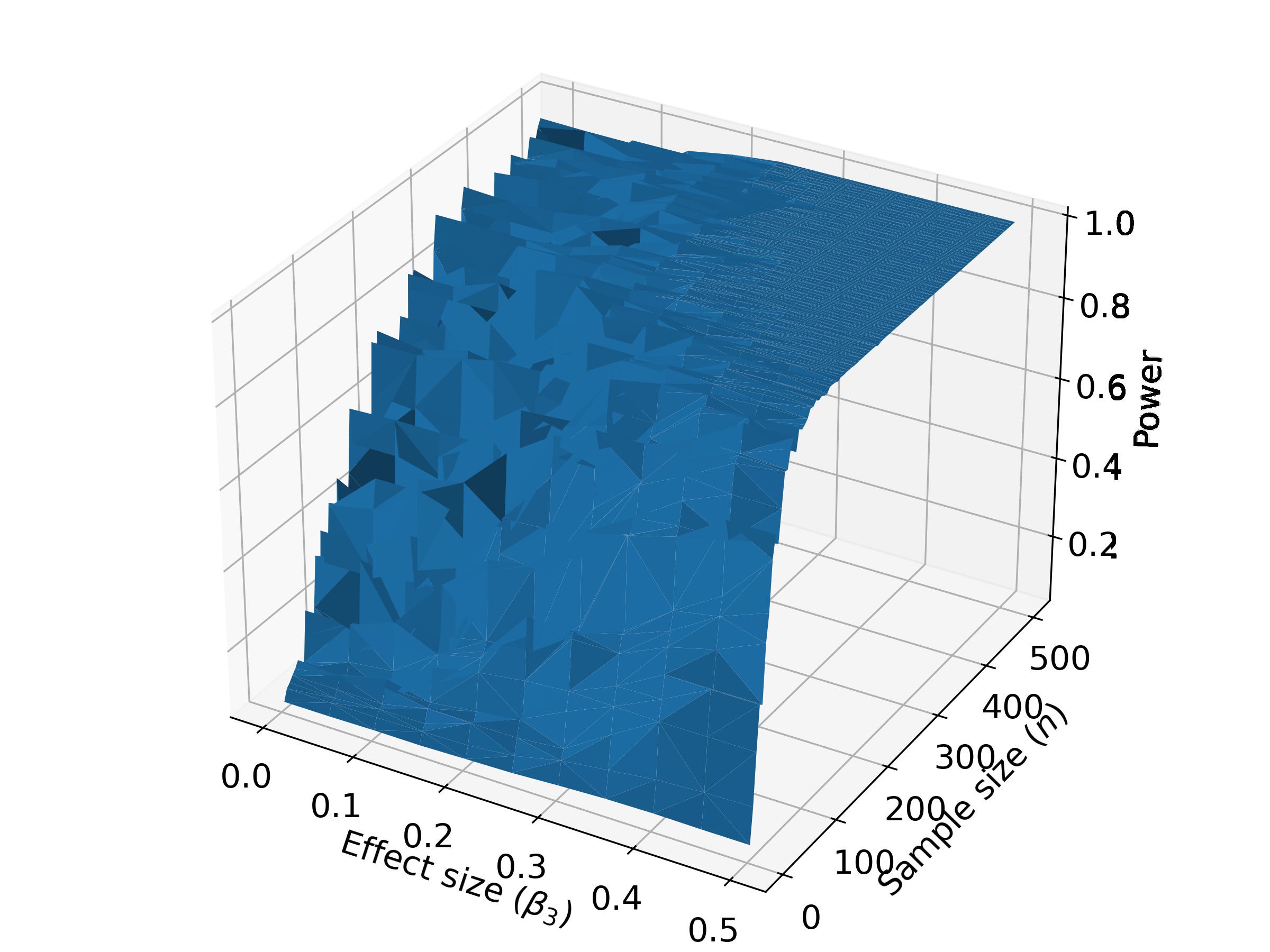}}
  \caption*{$I$ = 10}
\end{minipage}
\begin{minipage}{.32\textwidth}
  \centering
  \resizebox{\textwidth}{!}{\includegraphics{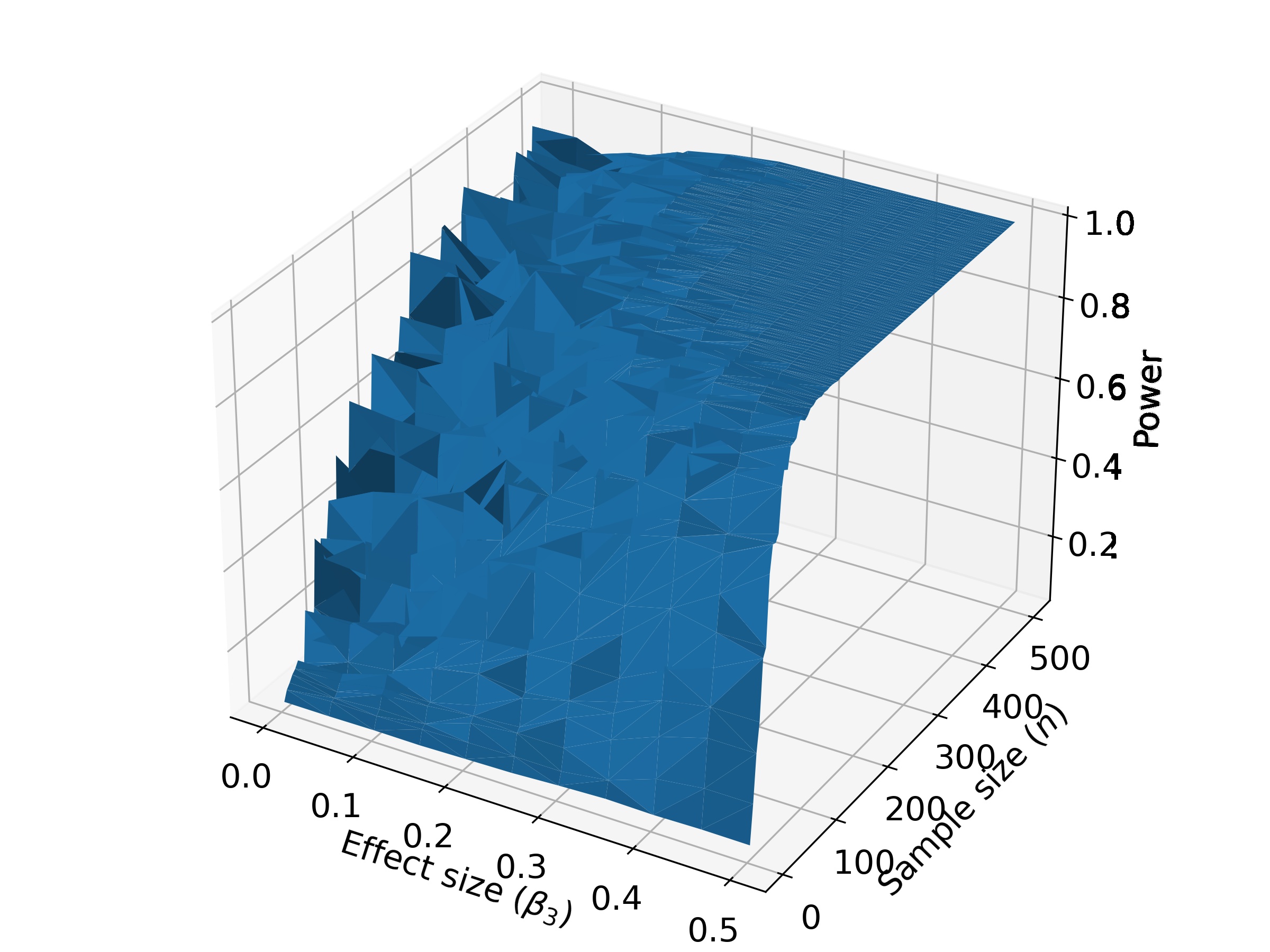}}
  \caption*{$I$ = 50}
\end{minipage}
\begin{minipage}{.33\textwidth}
  \centering
  \resizebox{\textwidth}{!}{\includegraphics{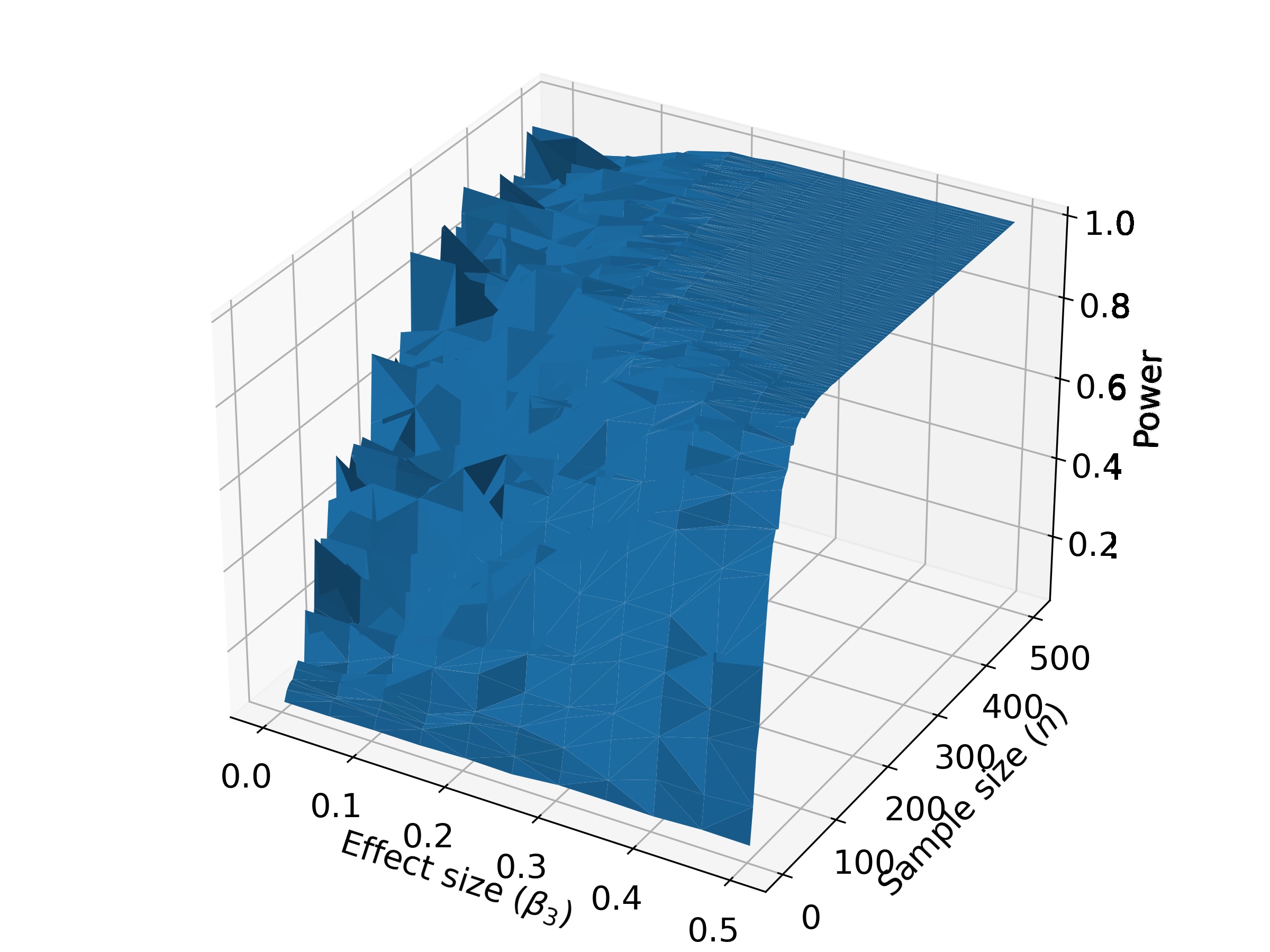}}
  \caption*{$I$ = 100}
\end{minipage}
\caption{Quality of learning the power manifold as a function of the effect size $\beta_3$ and sample size $n$ vs. the number of iterations $I$ when the population when $N=5000$.}  \label{fig:qualityoflearning35000}
\end{figure*}

%% file: sections/conclusion.tex
\section{Conclusion}
For learning the statistical power manifold in a time-efficient manner, we developed a novel genetic algorithm-based framework. Using our algorithm, applied researchers may learn/construct the statistical power manifold for some given initial constraints on different parameters. The learned surface can be used to identify high-power and low-power regions in the power manifold that can help applied researchers design their experiments better.

We performed experiments to demonstrate the performance of the proposed algorithm. We showed that our algorithm learns the power manifold as good as the costly brute-force methods while bringing huge savings in terms of run-time. Furthermore, we showed that the quality of learning using the proposed method improves as the number of iterations of the genetic algorithm increases. However, the run-time of the proposed algorithm also increases as the number of iterations of the genetic algorithm increases. This exhibits the optimality vs. run-time trade-off associated with our algorithm. Based on the estimation threshold and the availability of computational resources, the user may choose the required number of iterations and the size of the population.

\section{Future Work}
In the future, we are interested in also performing the genetic algorithm step of our algorithm for power minimization which would help us better learn the low-power regions. We are also interested in reducing the root mean squared error further by exploring more sophisticated prediction methods like neural networks. Furthermore, we are interested in extending our algorithm for learning the gradient of the statistical power manifold for different parameters. 
